\documentclass[%
 aip,
 amsmath,amssymb,
 preprint,%
]{revtex4-1}

\usepackage{graphicx}
\usepackage{dcolumn}
\usepackage{bm}
\usepackage{mathtools}
\usepackage{siunitx}
\usepackage{xcolor}
\usepackage{tikz}
\usetikzlibrary{shapes, arrows}
\tikzstyle{block} = [draw, fill=white, rectangle, minimum height=3em, minimum width=5em]
\tikzstyle{sum} = [draw, fill=white, circle, node distance=1cm]
\tikzstyle{input} = [coordinate]
\tikzstyle{output} = [coordinate]
\tikzstyle{pinstyle} = [pin edge={to-,thin,black}]

\begin{document}


\title[Synchronization of Chaotic Oscillators With Partial Linear Feedback Control]{Synchronization of Chaotic Oscillators With Partial Linear Feedback Control}

\author{K. Mistry}
\author{S. Dash}%
\altaffiliation{ 
B.Tech. (Electronics and Communication Engineering) student at National Institute of Technology (NIT), Trichy, India.
}%

\author{S. Tallur}
 \homepage{http://www.ee.iitb.ac.in/~stallur/index.php/}
\affiliation{%
Indian Institute of Technology (IIT) Bombay, Mumbai, India
}%
\email{stallur@ee.iitb.ac.in}

\date{\today}

\begin{abstract}
We present a methodology for synchronization of chaotic oscillators with linear feedback control. The proposed method is based on analyzing the chaotic oscillator as a multi-mode linear system and deriving sufficient conditions for asymptotic stability. The oscillators are synchronized in a master-slave configuration, wherein a subset of the state variables for implementing the feedback control, enabling applications in cryptography for message encryption using the unused chaotic state variables. Controller stability is ensured through conventional root-locus technique for designing appropriate loop gain. We validate the methodology presented here with numerical simulations and experimental results obtained using an operational amplifier (op-amp) based electronic chaotic oscillator circuit.
\end{abstract}

\pacs{Valid PACS appear here}
\keywords{Chaos; synchronization; partial linear feedback control; multi-mode system}
\maketitle

\section{\label{sec:level1}Introduction}
All second order dynamical systems exhibit one of three categories of trajectory in state space\cite{ref1}: 1) stable (convergent) 2) unstable (divergent) and 3) limit cycle (oscillatory). Higher order dynamical systems may exhibit another type of trajectory, namely chaotic behavior \cite{ref2,ref3,ref4,ref5,ref6}. Such systems may be emulated through simple electronic circuits \cite{ref7,ref8,ref9,ref10,ref11,ref12}, that exhibit rich non-linear dynamics while appearing deceptively deterministic from a circuit analysis perspective. Synchronization of chaotic oscillator circuits can enable several interesting applications in electronic message encryption\cite{ref13a,ref13,ref14}. Numerous methods for synchronization of chaotic systems have been proposed over the decades \cite{ref16,ref17,ref18b,ref18,ref19,ref20,ref21,ref22,ref23,ref24,ref25,ref26,ref27,ref28,ref29}, however all such implementations require either all state variables of the individual oscillators to generate the necessary locking signal to entrain the slave oscillators to the master oscillator \cite{ref16,ref17,ref18,ref19,ref20,ref21,ref22,ref23,ref24,ref25}, or a non-linear feedback signal employing a subset of state variables \cite{ref26,ref27,ref28,ref29}.

In this work we report a methodology to design a linear feedback controller to synchronize two chaotic oscillators represented by third order non-linear differential equations. The oscillators are analyzed as piecewise linear systems in different modes of operation. Using linear control theory and root locus method, the controller coefficients can be appropriately designed to ensure stability across all modes of operation, and utilizing a partial subset of state variables to generate the feedback signal. The unused state variables can then be employed for message encryption by adding these to a small-amplitude message signal at the transmitter in a communication system. The encrypted message could then be recovered at the receiver end by synchronizing the local oscillator at the receiver end to the transmitter oscillator, and subtracting the corresponding states used in encryption. We present a proof for the stability of this technique and provide validation with Scilab simulations of a third order non-linear system and experimental measurements obtained through an operational amplifier (op-amp) circuit implementation of the oscillators and the controller.

The paper is structured as follows: section II describes the notations and section III introduces the chaotic oscillator circuit used in this work. Section IV introduces some control systems techniques for synchronization, along with their limitations. Section V describes the method presented in this work in detail and a methodology for designing the controller, and section VI presents numerical simulations and experimental results corroborating this method.

\section{\label{sec:level1}Notations}

This section introduces the notations we use to describe the system mathematically. We focus on a chaotic oscillator represented by a third order non-linear differential equation:
\begin{equation}
    \frac{d^3x}{dt^3} = c\frac{d^2x}{dt^2} + b\frac{dx}{dt} + f(x).
    \label{eq1}
\end{equation}
where, $f(x)$ is piecewise linear function that captures the non-linearity in the system. We choose the following form of $f(x)$: 
\begin{equation}
    f(x) = 
    \begin{cases} 
      ax+u_1 & x<0 \\
      u_0 & x\geq 0 
   \end{cases}
   \label{eq2}
\end{equation}

Here $a,b,c,u_0$ and $u_1$ are all real constants. Defining state variables $x_1=x$, $x_2=\frac{dx}{dt}$ and $x_3=\frac{d^2x}{dt^2}$, state space realization of equation \ref{eq1} is expressed as follows:
\begin{equation}
    \begin{array}{l}
       \dot{x_1}=x_2\\
       \dot{x_2}=x_3\\
       \dot{x_3}=f(x_1)+bx_2+cx_3=g(x_1,x_2,x_3)
    \end{array}
    \label{eq3}
\end{equation}

The state vector for this state space model is expressed as $X=\begin{bmatrix}x_1 &x_2 &x_3\\\end{bmatrix}^T$. For synchronization of oscillators, we introduce a control signal to dictate the dynamics of the slave oscillator. The control signal is modeled as signal $u(t)$, and the combined model is expressed below:
\begin{equation}
    \frac{d^3x}{dt^3} = c\frac{d^2x}{dt^2} + b\frac{dx}{dt} + f(x) + u(t).
\end{equation}
\begin{equation}
    \begin{array}{l}
       \dot{x_1}=x_2\\
       \dot{x_2}=x_3\\
       \dot{x_3}=g(x_1,x_2,x_3) + u(t)
    \end{array}
\end{equation}


When two oscillators are synchronized, the trajectory in state-space is identical for both oscillators. We consider a master-slave locking scheme for two oscillators and denote the state space variables of the master oscillator as $x_i$ and those of the slave oscillator as $y_i$, $i=1,2,3$. The slave oscillator dynamics are also controlled through the controller output $u(t)$. The state space representation of both oscillators are then written as below:

\begin{equation}
    \begin{array}{l}
       \dot{x_1}=x_2\\
       \dot{x_2}=x_3\\
       \dot{x_3}=g(x_1,x_2,x_3),
    \end{array}
    \quad
    \begin{array}{l}
       \dot{y_1}=y_2\\
       \dot{y_2}=y_3\\
       \dot{y_3}=g(y_1,y_2,y_3)+u(t).
    \end{array}
    \label{eq6}
\end{equation}

The system is easier to analyze in terms of the error states:  $e_i=y_i-x_i$, $i=1, 2, 3$. Synchronization of the two oscillators requires that the states $e_1$, $e_2$ and $e_3$ converge to zero. From equations \ref{eq3} and \ref{eq6}, we obtain:

\begin{equation}
    \begin{array}{l}
       \dot{e_1}=e_2\\
       \dot{e_2}=e_3\\
       \dot{e_3}=f(y_1)-f(x_1)+be_2+ce_3+u(t)
    \end{array}
    \label{eq7}
\end{equation}

The error states can be expressed as a vector $E=\begin{bmatrix}e_1 &e_2 &e_3\\\end{bmatrix}^T$.

\section{Circuit implementation of the chaotic oscillator}

\begin{figure}[!htb]
	\begin{center}
		\includegraphics[width=1\columnwidth]{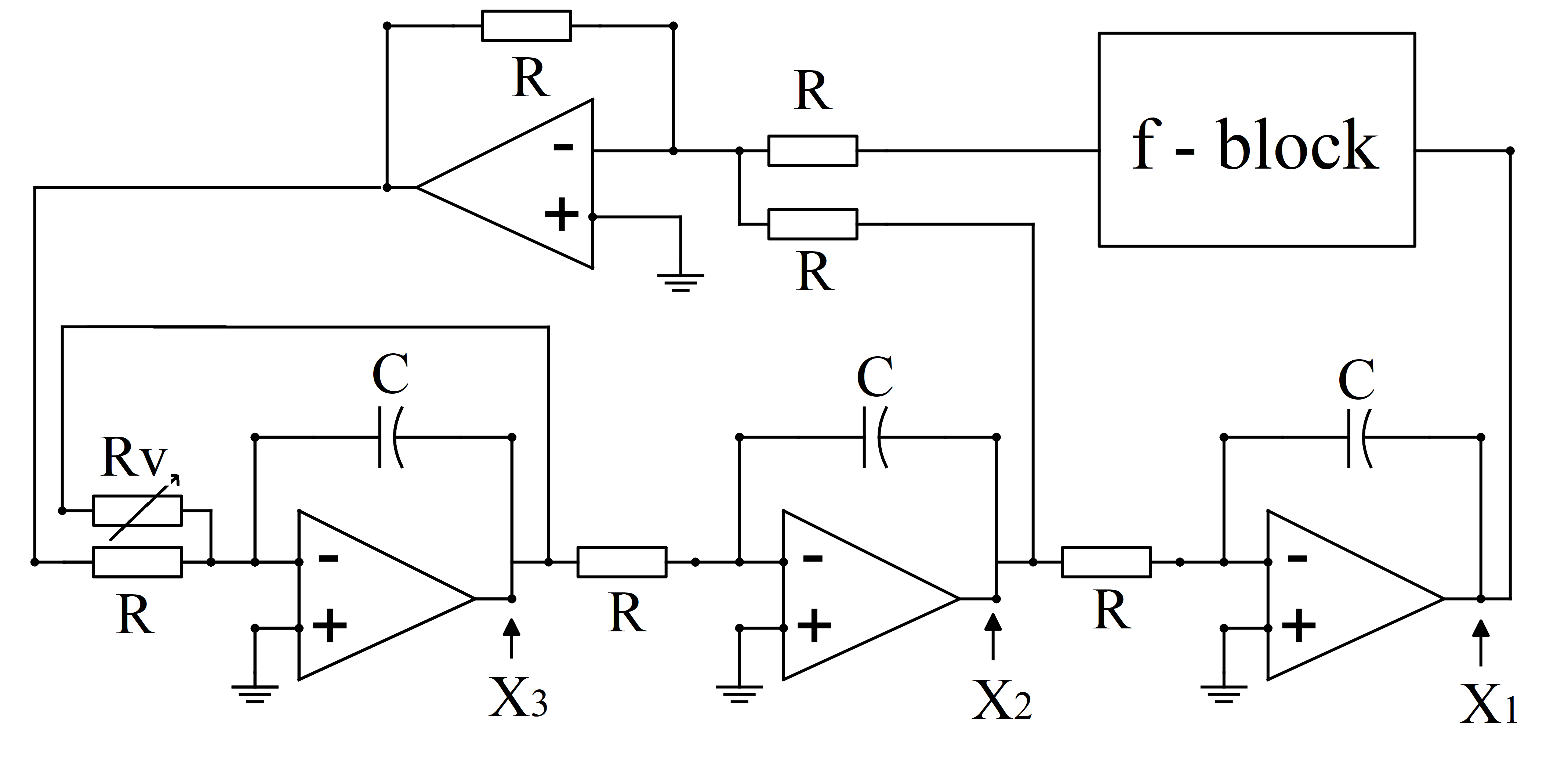}
	\end{center}
	\caption{Circuit diagram of the chaotic signal generator (oscillator), based on an architecture proposed by Kiers et al. \cite{ref7}. The ``f-block'' is a non-linear circuit shown in Figure \ref{fblockfig}.}
	\label{cktdiag}
\end{figure}

\begin{figure}[!htb]
	\begin{center}
		\includegraphics[width=1\columnwidth]{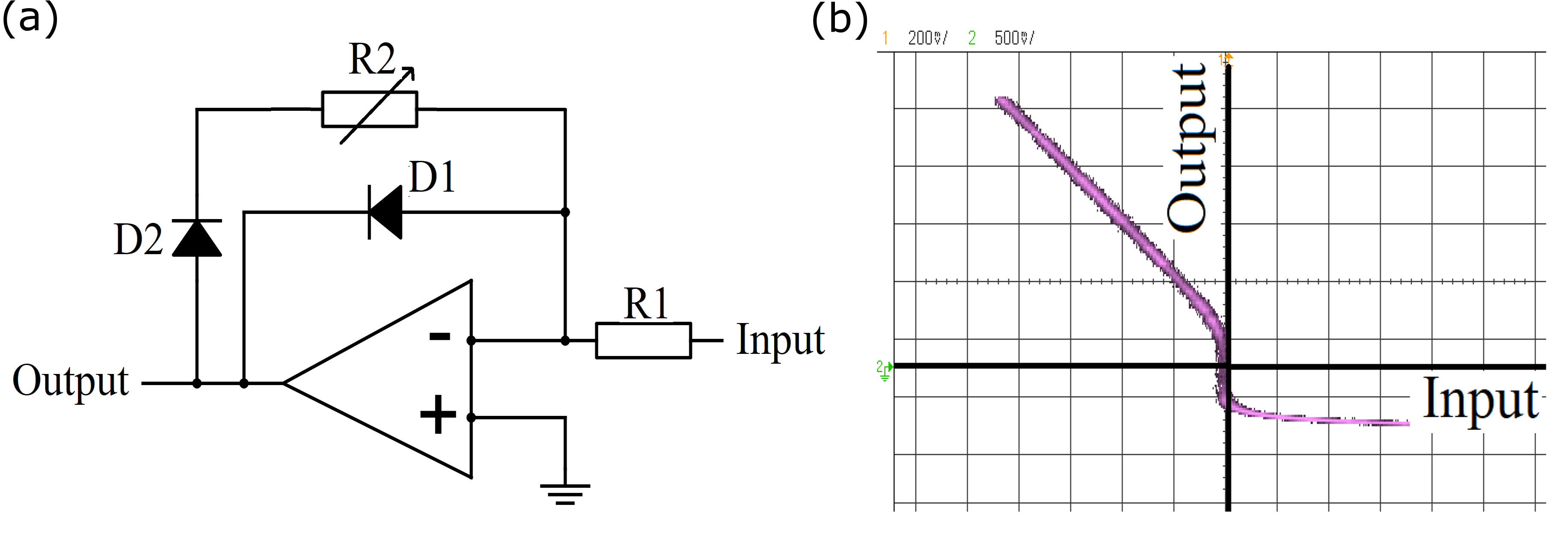}
	\end{center}
	\caption{(a) Circuit diagram for the ``f-block'' in Figure \ref{cktdiag}, that implements the equation for $f(x)$ as in equation (\ref{eq11}). (b) Experimentally measured transfer function of the non-linear f-block, verified by applying sinusoidal signal to the f-block circuit and observing output vs input graph on oscilloscope (configured to display in XY mode).}
	\label{fblockfig}
\end{figure}

For experimental validation of the technique, we implement the system differential equation (\ref{eq1}) using an analog circuit containing resistors, capacitors, operational amplifiers (op-amps) and diodes as shown in Figure \ref{cktdiag}. The chaotic behavior of such circuits has been extensively studied and documented by Kiers et al.\cite{ref7}. The difference in this circuit is the implementation of the ``f-block'' as shown in Figure \ref{fblockfig}, which implements a modified precision rectifier circuit. The characteristic differential equation of this circuit is expressed as: 

\begin{equation}
    \frac{d^3x}{dt^3} = - \frac{1}{R_vC} \frac{d^2x}{dt^2} - \frac{1}{R^2C^2} \frac{dx}{dt} + \frac{1}{R^3C^3} f(x)
    \label{eq10}
\end{equation}


The f-block circuit in Figure \ref{fblockfig} implements the following function:

\begin{equation}
    f(x) = 
    \begin{cases} 
      -\frac{R_2}{R_1}x+0.7 & x<0 \\
      -0.7 & x\geq 0 
   \end{cases}
   \label{eq11}
\end{equation}

Notice that for values of $x\geq 0$, the function $f(x)$ has a non-zero value due to the forward bias voltage drop across diode $D1$. This modification does away with the requirement of an external bias voltage that is necessary in the implementation reported by Kiers et al. \cite{ref7}. 

\section{\label{contrl_tech}Control systems techniques for synchronization}


Since the system under consideration is governed by a non-linear transfer function, several non-linear control techniques \cite{ref27,ref28,ref29,ref30} can be used to control the dynamics and achieve synchronization of the two oscillators. Consider feedback linearization technique \cite{ref1} applied to this system, wherein we design $u(t)$ such that the overall system becomes linear in nature. Observing equation (\ref{eq7}) we can select $u(t)=-f(y_1)+f(x_1)+v(t)$. The state space representation of the error states can then be rewritten as follows:

\begin{equation}
    \begin{array}{l}
       \dot{e_1}=e_2\\
       \dot{e_2}=e_3\\
       \dot{e_3}=be_2+ce_3+v(t)
    \end{array}
    \label{feedlin}
\end{equation}

\begin{equation}
       \dot{E}=\begin{bmatrix}
    0  & 1 & 0\\
    0  & 0 & 1\\
    0  & b & c\\
    \end{bmatrix} E
    +
    \begin{bmatrix}
    0\\
    0\\
    1\\ 
    \end{bmatrix}v(t)
    \label{feedlins}
\end{equation}

As evident from equations (\ref{feedlin}) and (\ref{feedlins}), the system is transformed to a linear system, with state space equation of form $\dot{E}=AE+Bv(t)$ as shown in equation (\ref{feedlins}). The signal $v(t)$ is chosen as a linear combination of the error states, i.e. $v(t)=KE$, such that the overall state matrix $A+BK$ is Hurwitz. This is a necessary and sufficient condition for stability of the controller, as will be explained in detail in section \ref{condition}.  In this technique, the controller implementation $u(t)$ depends on the non-linearity in the system $f(x)$. Even though one may discretely implement a controller by externally implementing the non-linearity, the technique is susceptible to drifts in the system that may change the nature of $f(x)$, and implementing such a robust controller may not be feasible practically.

Another method to design the controller is by approximating the non-linearity in the system transfer function as a smooth (continuous and differentiable) response e.g. as a higher order polynomial. However this approximation is effective only in the vicinity of equilibrium point(s) of the system (in this case origin) i.e. the errors are low for small signal amplitudes\cite{ref1}. This scheme is not robust as large signals at any of the circuit nodes at start-up (initial conditions) will lead to large diverging errors and the controller may not achieve synchronization.

\section{\label{sec:level1}Analysis as Multi-mode linear system}

The dynamics of a chaotic systems may also be viewed as a trajectory switching across various modes, and studied as a Linear Complimentarity System (LCS)\cite{ref30}. The system under consideration can be expressed in LCS form as follows:

\begin{equation}
    \dot{X}(t) = C_1X(t) + C_2w'(t) + C_3u'(t)
\end{equation}
\begin{equation}
    y'(t) = C_4X(t) + C_5w'(t) + C_6u'(t)
\end{equation}

where $C_i$ $(i = 1,2,\dots,6)$ are matrices of appropriate size, $\dot X=\begin{bmatrix}\dot x_1 &\dot x_2 &\dot x_3\\\end{bmatrix}^T$, $X=\begin{bmatrix}x_1 &x_2 &x_3\\\end{bmatrix}^T$ and $u'(t) \geq 0, y'(t) \geq 0, u'(t)^T y'(t)=0$. For the oscillator circuit, this translates to:

\begin{equation}
    \dot{X}(t) = \begin{bmatrix}
    0  & 1 & 0\\
    0  & 0 & 1\\
    0 & \frac{-1}{R^2C^2} & \frac{-1}{R_vC}\\
    \end{bmatrix} X(t)
    +
    \begin{bmatrix}
    0\\
    0\\
    - \frac{0.7}{R^3C^3}\\ 
    \end{bmatrix}
    +
    \begin{bmatrix}
    0\\
    0\\
    1\\ 
    \end{bmatrix}u'(t)
\end{equation}
\begin{equation}
    y'(t) = \begin{bmatrix}
    \frac{1}{R^3C^3} \frac{R_2}{R_1} & 0 & 0\\
    \end{bmatrix} X(t)
    - 2\frac{0.7}{R^3C^3} +u'(t)
\end{equation}

The system input is denoted as $w'(t)$ and the switching vectors in the system, i.e. $u'(t)$ and $y'(t)$, evolve such that one of them will be zero and other will be non-negative at every instant in time \cite{ref30}. If $u'(t)=0$, we obtain state space equation with constraint $x_1 \geq 0$ and if $y'(t)=0$ we obtain another state space equation with constraint $x_1 < 0$. While one may use stability theories for LCS \cite{ref30} to design a suitable controller, a more intuitive approach is to analyze the system as a multi-mode linear system and study stability of each mode using standard linear control theory. This technique forms the heart of the work presented here, and is described in detail below:



\subsection{Multi-mode representation of the control system}

The piecewise linear function $f(x)$ appears in equation (\ref{eq7}), and hence the system shows four modes of operation, depending on the signs of $x_1$ and $y_1$:

$\:$

\textbf{MODE-I ($x_1\geq 0 \quad y_1 \geq 0$):}

\begin{center}
    $
       \dot{E}=\begin{bmatrix}
    0  & 1 & 0\\
    0  & 0 & 1\\
    0  & b & c\\
    \end{bmatrix} E
    +
    \begin{bmatrix}
    0\\
    0\\
    1\\ 
    \end{bmatrix}u(t)
    $
\end{center}

\textbf{MODE-II  ($x_1 < 0 \quad y_1 \geq 0$):} 

\begin{center}
    $
       \dot{E}=\begin{bmatrix}
    0  & 1 & 0\\
    0  & 0 & 1\\
    0  & b & c\\
    \end{bmatrix} E
    +
    \begin{bmatrix}
    0\\
    0\\
    u_0-ax_1-u_1\\ 
    \end{bmatrix}
    +
    \begin{bmatrix}
    0\\
    0\\
    1\\ 
    \end{bmatrix}u(t)
    $
\end{center}

\textbf{MODE-III  ($x_1 < 0 \quad y_1 < 0$):} 

\begin{center}
    $
       \dot{E}=\begin{bmatrix}
    0  & 1 & 0\\
    0  & 0 & 1\\
    a  & b & c\\
    \end{bmatrix} E
    +
    \begin{bmatrix}
    0\\
    0\\
    1\\ 
    \end{bmatrix}u(t)
    $
\end{center}

\textbf{MODE-IV  ($x_1 \geq 0 \quad y_1 < 0$):} 

\begin{center}
    $
       \dot{E}=\begin{bmatrix}
    0  & 1 & 0\\
    0  & 0 & 1\\
    0  & b & c\\
    \end{bmatrix} E
    +
    \begin{bmatrix}
    0\\
    0\\
    ay_1+u_1-u_0\\ 
    \end{bmatrix}
    +
    \begin{bmatrix}
    0\\
    0\\
    1\\ 
    \end{bmatrix}u(t)
    $
\end{center}

The equation in mode IV can be rewritten as below, by writing $ay_1 = ae_1 + x_1$:

\begin{center}
    $
       \dot{E}=\begin{bmatrix}
    0  & 1 & 0\\
    0  & 0 & 1\\
    a  & b & c\\
    \end{bmatrix} E
    +
    \begin{bmatrix}
    0\\
    0\\
    ax_1+u_1-u_0\\ 
    \end{bmatrix}
    +
    \begin{bmatrix}
    0\\
    0\\
    1\\ 
    \end{bmatrix}u(t)
    $
\end{center}

\subsection{Conditions for stability of controller}\label{condition}
For any linear autonomous system $\dot{X}=AX$, the matrix $A$ is called state matrix of the system, and its eigenvalues are the poles of the system transfer function. The eigenvalues of matrix $A$ are the roots of its characteristic polynomial, $\Delta_{A}(s) = det(sI-A)$. Matrix $A$ is called a Hurwitz matrix if all roots of $\Delta_{A}(s)$ lie in the left half of the complex plane, i.e. all roots have strictly negative real part. Consequently a linear system is asymptotically stable at origin if it has a Hurwitz state matrix \cite{ref31}. For a linear system $\dot{X}=AX+Bu$, if matrix $A$ is Hurwitz then system is BIBO (bounded input bounded output) stable, i.e. if the values of the input to the system are bounded, the output of the system also necessarily has bounded range of values \cite{ref31}. A bounded signal in this context refers to a signal that has finite magnitude at every instance in time.

In each of these modes, the coupled oscillators are described by a linear system of equations. To stabilize such a system, $u(t)$ may also be a designed as a linear feedback controller. Let us denote $u(t)= KE$, where $K=\begin{bmatrix}\alpha &\beta &\gamma\\\end{bmatrix}$. Hence $u(t)=\alpha e_1+\beta e_2+\gamma e_3$. The state matrix for modes I and II is rewritten as $A_0=\begin{bmatrix}
    0  & 1 & 0\\
    0  & 0 & 1\\
    \alpha  & b' & c'\\
    \end{bmatrix}$, where $b'=b+\beta$ and $c'=c+\gamma$. The state matrix for modes III and IV is rewritten as: $A_1=\begin{bmatrix}
    0  & 1 & 0\\
    0  & 0 & 1\\
    a'  & b' & c'\\
    \end{bmatrix}$, where $a'=a+\alpha$.

The controller coefficients $\alpha$, $\beta$ and $\gamma$ can be tuned to ensure that all eigenvalues of $A_0$ and $A_1$ lie in the left half of complex plane, and consequently the system is asymptotically stable at origin for modes I and III. For modes II and IV, the state matrix is Hurwitz, and hence the system is BIBO stable. The system equation in these modes also contains an input term proportional to state $x_1$. Since $x_1$ is a state variable of the master chaotic oscillator (implemented as an op-amp based electronic circuit), its magnitude is bounded. Thus the error state variables $e_1$, $e_2$ and $e_3$ are also bounded in modes II and IV.

As the trajectory of the error state variable system evolves in time in state space, it switches from one mode to another. Notice that if the system trajectory enters mode II or mode IV, the following conditions are always true: i) the magnitude of the error state trajectory remains bounded due to BIBO stability of the system, and ii) the trajectory can evolve to another mode as the magnitude and sign of the state $x_1$ independently changes with time. In modes I and III, the trajectory of the error state variable system asymptotically converges to origin. Designing $A_0$ and $A_1$ matrices to be Hurwitz thus stabilizes the controller, and ensures that state vector $E$ will converge to origin, i.e. the two chaotic oscillators will synchronize. It is worth noting that the individual stability of each mode is a sufficient, but not a necessary condition for ensuring synchronization. If the rate of increment in distance of the state trajectory point from origin (divergence) in the unstable modes is lower than the rate of decrement in distance of state trajectory point from origin (convergence) in a stable mode, the overall state trajectory of the multi-mode system will eventually converge towards origin.

\subsection{Design of controller using root-locus approach}\label{contr_des}

The controller $u(t)$ is constructed as a linear combination of all error states $e_i$, $i=1,2,3$. In some applications all states are either not available or cannot be used for constructing the controller, e.g. in cryptography applications, where one or more states may be required for message encryption, and remaining states are used to construct the controller to synchronize the receiver oscillator to the transmitter oscillator for message decryption. In such cases, $u(t)$ can simply be a scaled version of any one of the states. Consider $u(t)=\alpha e_1$, and values of $\beta$ and $\gamma$ will be zero. The characteristic polynomial of matrix $A_0$ is thus:
\begin{equation}
\Delta_{A_0}(s) = s^3 - c s^2 - b s - \alpha
\label{eq8}
\end{equation}

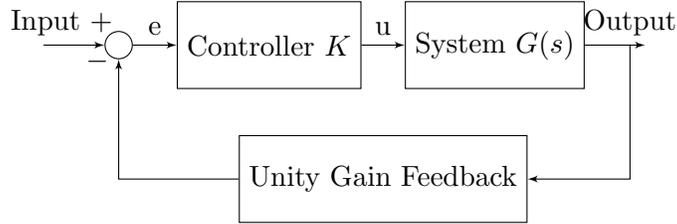
\begin{figure}
\centering

\begin{tikzpicture}[auto, node distance=2cm,>=latex']
    \node [input, name=input] {};
    \node [sum, right of=input] (sum) {};
    \node [block, right of=sum, node distance=2cm] (controller) {Controller $K$};
    \node [block, right of=controller, node distance=3cm] (system) {System  $G(s)$};
    \draw [->] (controller) -- node[name=u] {u} (system);
    \node [output, right of=system] (output) {};
    \node [block, below of=u] (feedback) {Unity Gain Feedback};

    \draw [draw,->] (input) -- node[pos=0.3] {Input \textbf{$+$}} 
        node [near end] {} (sum);
    \draw [->] (sum) -- node {e} (controller);
    \draw [->] (system) -- node [name=y, xshift=0.2cm] {Output}(output);
    \draw [->] (y) |- (feedback);
    \draw [->] (feedback) -| node[pos=0.99] {\textbf{$-$}} 
        node [near end] {} (sum);
\end{tikzpicture}
\caption{Generalized representation of a control loop for root locus analysis. The root locus technique is used to design a stable linear controller for synchronization of the two oscillators.}
\label{bd1}
\end{figure}

To analyze how the roots of this polynomial vary with value of $\alpha$, we use root locus analysis. The root locus plot of any system graphically illustrates the trajectory of variation of the roots of the system characteristic equation in the complex plane, when some parameter of the system is varied \cite{ref31}. Consider a system with transfer function $G(s)$ controlled using negative unity gain feedback and proportional controller with gain $K$ as shown in Figure \ref{bd1}. The closed loop transfer function is given by $T(s)=\frac{KG(s)}{1+KG(s)}$ and the characteristic polynomial $\Delta(s)$ of this closed loop system is the denominator in $T(s)$. The root locus of this system is a plot of the roots of $\Delta(s)$ in the complex plane as $K$ is varied from 0 to $\infty$. Now consider a system with open loop transfer function $G_{A_0}(s)$ as given in equation (\ref{eq9}) and proportional controller gain $K=-\alpha$.

\begin{equation}
G_{A_0}(s) = \frac{1}{s^3 - c s^2 - b s}
\label{eq9}
\end{equation}

The characteristic polynomial of this closed loop system is $\Delta_{A_0}(s)$ as expressed in equation (\ref{eq8}). We can choose suitable value of $\alpha$ by examining the root locus of $G_{A_0}(s)$ such that all roots of $\Delta_{A_0}(s)$ lie in the left-half of the complex plane (i.e. the real part of the roots are all negative), thus ensuring that matrix $A_0$ is Hurwitz. Following a similar procedure with a suitably designed $G(s)$, we can find suitable values of $\alpha$ such that matrix $A_1$ is also Hurwitz. If no such values of $\alpha$ can be identified, we can instead try $u(t)=\beta e_2$ or $u(t)=\gamma e_3$ and repeat the same root locus exercise. If one error state alone proves insufficient to generate a stable controller, one can then explore using a linear combination of multiple states for this exercise, depending on how many states are available for controller design based on the application.

\section{Simulation and Experimental results}

The oscillator circuit and the f-block shown in Figure \ref{cktdiag} and Figure \ref{fblockfig}(a) respectively are implemented using variable resistors $R_v$ and $R_2$. The fixed resistance values are $R=47k\Omega$, $R_1=10k\Omega$ and the variable resistors are tuned to operate the oscillator in the chaotic regime. All capacitors are implemented as ceramic capacitors with capacitance $C=0.1nF$, and the op-amps are implemented using IC TL071 low-noise JFET-input general-purpose operational amplifier ICs from Texas Instruments. The diodes in Figure \ref{fblockfig}(a) are implemented using 1N4148 silicon diodes. The dynamics of the system are simulated by solving the differential equation numerically in Scilab. In our simulation we modify equation (\ref{eq10}) by scaling time as $t=(RC)*T$, to obtain the modified differential equation expressed in equation (\ref{eq12}). Comparing equations (\ref{eq12}) and (\ref{eq11}) with equations (\ref{eq1}) and (\ref{eq2}) respectively, we obtain $b=-1$, $u_1=0.7$ and $u_0=-0.7$.

\begin{equation}
    \frac{d^3x}{dT^3} = - \frac{R}{R_v} \frac{d^2x}{dT^2} - \frac{dx}{dt} + f(x).
    \label{eq12}
\end{equation}

\begin{figure}[!htb]
	\begin{center}
		\includegraphics[width=0.8\columnwidth]{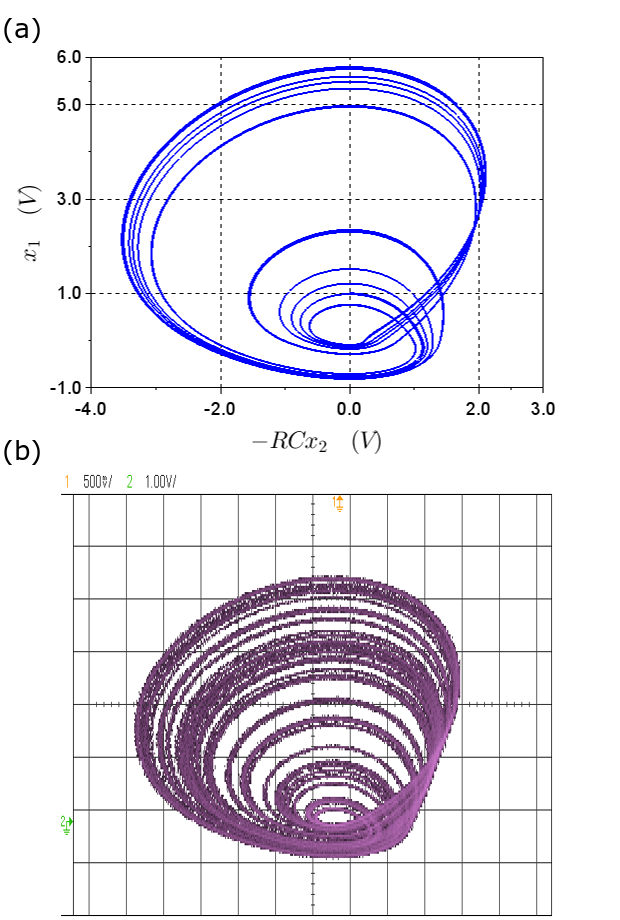}
	\end{center}
	\caption{(a) Simulated phase portrait of the oscillator described in equation (\ref{eq12}), obtained using numerical simulation in Scilab. (b) Experimentally measured phase portrait of the oscillator on an oscilloscope configured to display in $XY$ mode. The Y-axis displays signal at the node $x_1$ and X-axis displays signal at node $x_2$.}
	\label{chaotic}
\end{figure}

This non-linear differential equation can be simulated with different values of $a$ and $c$ to find the appropriate set of values to operate the oscillator in chaotic regime. Figure \ref{chaotic}(a) and Figure \ref{chaotic}(b) show the simulated and experimentally measured phase portrait of the oscillator using $R_v=71.1k\Omega$ $(c=-0.66)$ and $R_2=58k\Omega$ $(a=-5.8)$, which confirm the chaotic behavior. The experimental measurements are obtained on a Keysight DSOX 2002A oscilloscope configured to display signals in the $XY$ mode. 

Two such circuits are constructed and the steps illustrated in section \ref{contr_des} are implemented to design a linear controller $u(t)=\beta e_2$ to synchronize the two chaotic circuits. The characteristic polynomial of matrix $A_0$ is given by:
\begin{equation}
\Delta_{A_0}(s) = s^3 + 0.66 s^2 + (1-\beta) s
\label{eq13}
\end{equation}

\begin{figure}
\centering

\begin{tikzpicture}[auto, node distance=2cm,>=latex']
    \node [input, name=input] {};
    \node [sum, right of=input] (sum) {};
    \node [block, right of=sum] (controller) {$K=1-\beta$};
    \node [block, right of=controller, node distance=3cm] (system) {$G(s)=\frac{1}{s^2 + 0.66s}$};
    \draw [->] (controller) -- node[name=u] {} (system);
    \node [output, right of=system, node distance=2.5cm] (output) {};
    \node [block, below of=u] (feedback) {Unity Gain Feedback};

    \draw [draw,->] (input) -- node[pos=0.8] {\textbf{$+$}} 
        node [near end] {} (sum);
    \draw [->] (sum) -- node {} (controller);
    \draw [->] (system) -- node [name=y, xshift=0.1cm] {}(output);
    \draw [->] (y) |- (feedback);
    \draw [->] (feedback) -| node[pos=0.99] {\textbf{$-$}} 
        node [near end] {} (sum);
\end{tikzpicture}
\caption{Block diagram of control loop for root locus analysis of characteristic polynomial of matrix $A_0$. The trajectory of the roots of the loop transfer function are analyzed in complex plane as the gain parameter $K$ is varied from $0$ to $+\infty$.}
\label{bd2}
\end{figure}
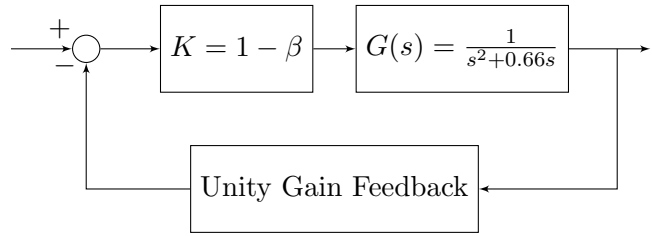

\begin{figure}[!htb]
	\begin{center}
		\includegraphics[width=0.6\columnwidth]{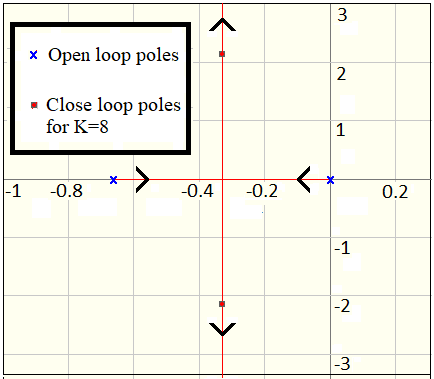}
	\end{center}
	\caption{Root locus for controller shown in Figure \ref{bd2}. The roots lie in the left half of the complex plane for all values of gain $K \geq 0$.}
	\label{rlocus1}
\end{figure}

To identify a suitable value of $\beta$ to ensure controller stability, we simulate the root locus of the control loop shown in Figure \ref{bd2}. Figure \ref{rlocus1} shows the root locus plot of the system shown in Figure \ref{bd2}, simulated using RootLocs\cite{ref32}, a freely distributed root locus plotting software. The roots always lie in the left-half of the complex plane for all values of $K$, and thus the system with state matrix $A_0$ will be asymptotically stable at origin for  $K \in [0,+\infty)$, i.e. $\beta \in (-\infty ,1]$.

A similar approach is employed to design matrix $A_1$ to be Hurwitz. The characteristic polynomial of matrix $A_1$ is expressed as:

\begin{equation}
\Delta_{A_1}(s) = s^3 + 0.66 s^2 + (1-\beta) s + 5.8
\label{eq15}
\end{equation}

\begin{figure}
\centering

\begin{tikzpicture}[auto, node distance=2cm,>=latex']
    \node [input, name=input] {};
    \node [sum, right of=input] (sum) {};
    \node [block, right of=sum] (controller) {$K=1-\beta$};
    \node [block, right of=controller, node distance=3cm] (system) {$G(s)=\frac{s}{s^3 + 0.66s^2 + 5.8}$};
    \draw [->] (controller) -- node[name=u] {} (system);
    \node [output, right of=system, node distance=2.5cm] (output) {};
    \node [block, below of=u] (feedback) {Unity Gain Feedback};

    \draw [draw,->] (input) -- node[pos=0.8] {\textbf{$+$}} 
        node [near end] {} (sum);
    \draw [->] (sum) -- node {} (controller);
    \draw [->] (system) -- node [name=y, xshift=0.1cm] {}(output);
    \draw [->] (y) |- (feedback);
    \draw [->] (feedback) -| node[pos=0.99] {\textbf{$-$}} 
        node [near end] {} (sum);
\end{tikzpicture}
\caption{Block diagram of control loop for root locus analysis of characteristic polynomial of matrix $A_1$. The trajectory of the roots of the loop transfer function are analyzed in complex plane as the gain parameter $K$ is varied from $0$ to $+\infty$.}
\label{bd3}
\end{figure}
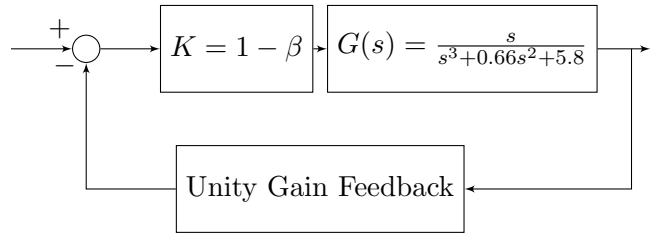

\begin{figure}[!htb]
	\begin{center}
		\includegraphics[width=0.6\columnwidth]{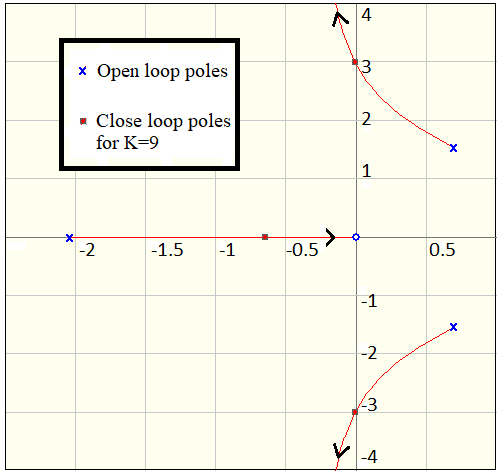}
	\end{center}
	\caption{Root locus for controller shown in Figure \ref{bd3}. The roots lie in the left half of the complex plane for values of gain $K \geq 9$.}
	\label{rlocus2}
\end{figure}

To observe the variation of roots of $\Delta_{A_1}(s)$ as we tune $\beta$, we study the root locus of closed loop system shown in Figure \ref{bd3}. Figure \ref{rlocus2} shows root locus plot of the system shown in Figure \ref{bd3}. Asymptotic stability of this system requires $K \in [9,+\infty)$, i.e. $\beta \in (-\infty ,-8]$. The system will be stable in all four modes when both matrices $A_0$ and $A_1$ are Hurwitz, i.e. when $\beta \in (-\infty ,-8]$.

\begin{figure}[!htb]
	\begin{center}
		\includegraphics[width=1\columnwidth]{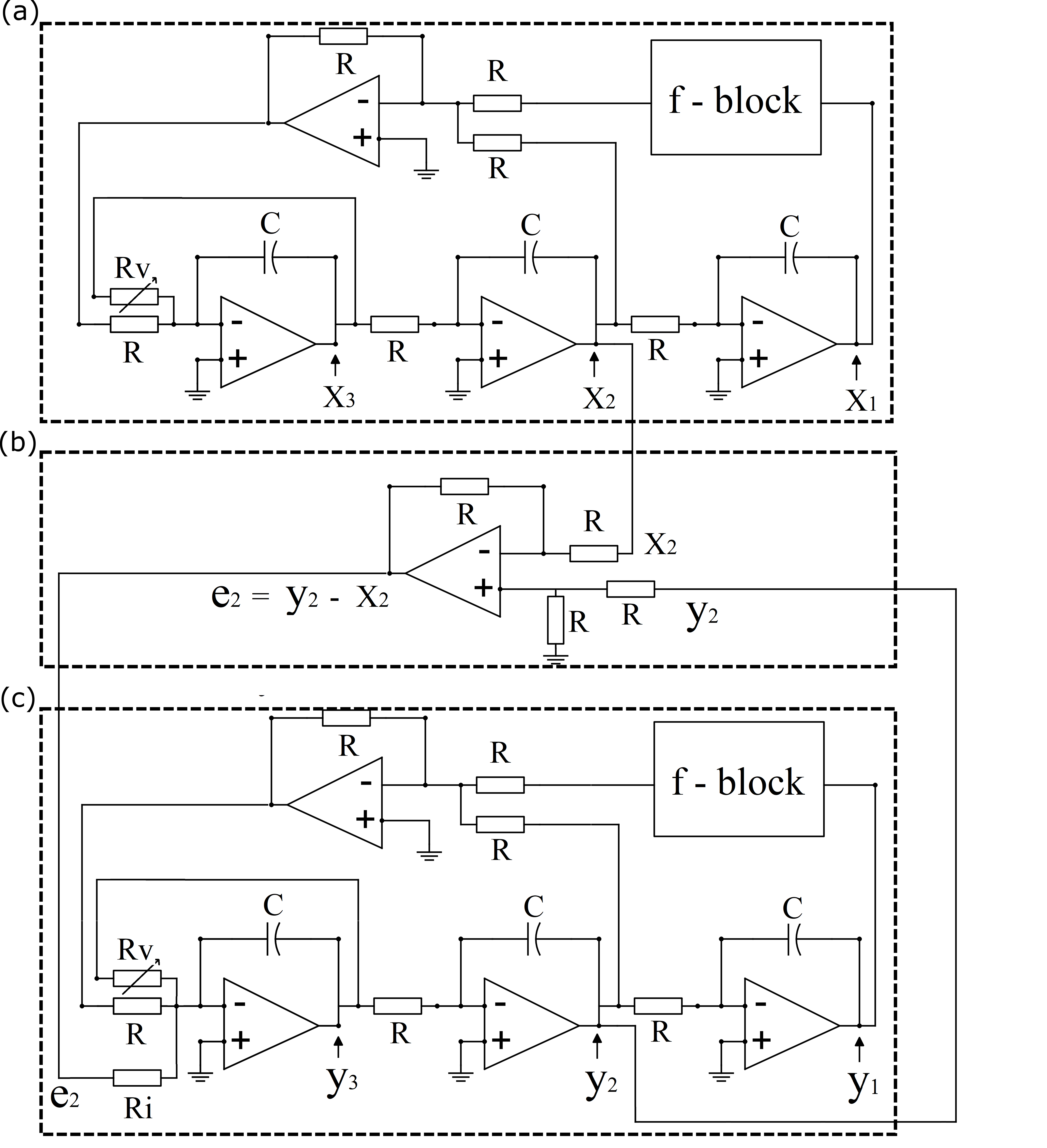}
	\end{center}
	\caption{Circuit diagram showing the (a) master chaotic circuit, (b) linear controller circuit, and (c) slave chaotic circuit with feedback controller: $u(t)=-\frac{R}{R_i} e_2(t)$}
	\label{cktdiagslave}
\end{figure}

Choosing any value of $\beta$ in this range allows us to design the controller as a signal proportional to $e_2=y_2-x_2$, where $y_2$ is the signal from the slave oscillator and $x_2$ is the signal from the master oscillator. The error signal $e_2$ is thereby generated using an unit gain op-amp differential amplifier with inputs $y_2$ and $x_2$, and is connected to the input of the slave oscillator circuit wherein it is scaled by gain $\beta = -\frac{R}{R_i}$. Figure \ref{cktdiagslave} shows the circuit diagram in its entirety.

\begin{figure}[!htb]
	\begin{center}
		\includegraphics[width=0.8\columnwidth]{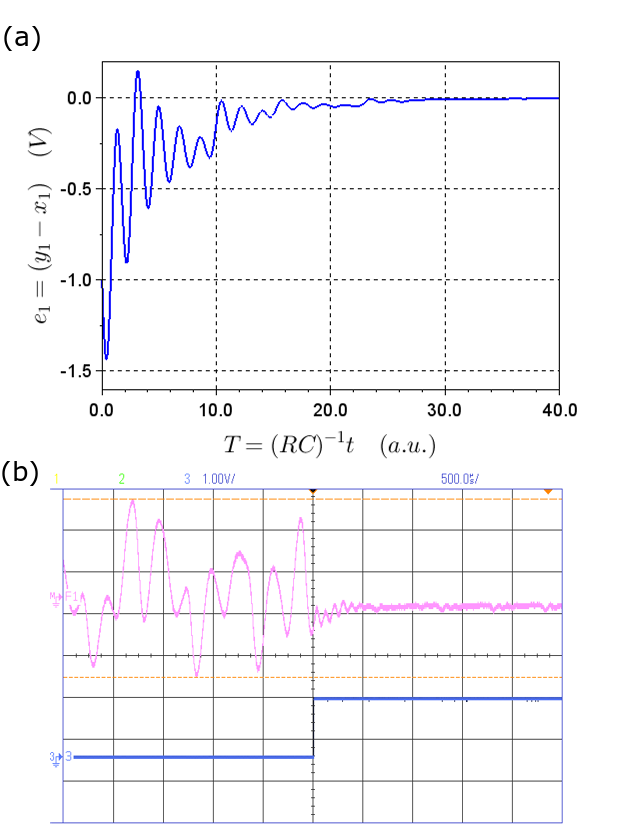}
	\end{center}
	\caption{Response time of controller: (a) Numerical simulation (in Scilab) shows the error state $e_1$ converging to zero after connecting the control signal $u(t)=-10e_2(t)$ at time $T = 0$. (b) Experimental result observed on an oscilloscope, wherein the error state $e_1$ converges to zero (upper trace) when the control signal is turned on using an electrical switch gated by the voltage step signal shown in the bottom trace.}
	\label{synch}
\end{figure}

\begin{figure}[!htb]
	\begin{center}
		\includegraphics[width=0.8\columnwidth]{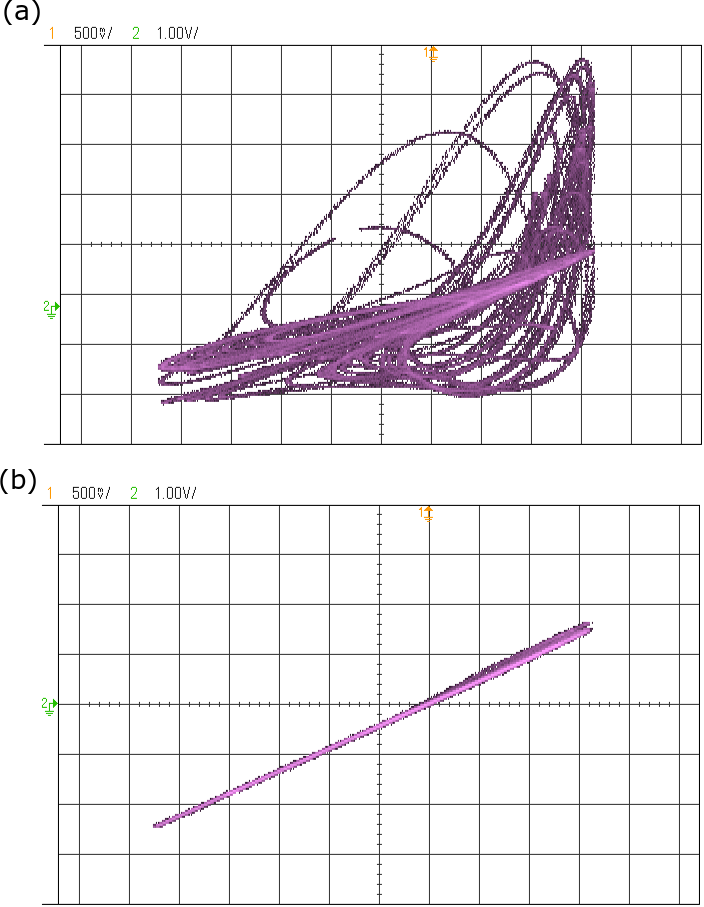}
	\end{center}
	\caption{Signals $x_1$ (master oscillator, Y-axis) and $y_1$ (slave oscillator, X-axis) observed on an oscilloscope configured to display in $XY$ mode. (a) In the unsynchronized state, the two signals are not correlated to each other. (b) When the two oscillators are synchronized, the two signals track each other and are equal in magnitude.}
	\label{synchxy}
\end{figure}

Root-locus analysis suggests that $-R/R_i = \beta \leq -8$ will ensure a stable controller and synchronization of the chaotic oscillators. Figure \ref{synch}(a) shows numerical simulation for the error state $e_1$ converging to zero when the controller is turned on at time $t=0$, for gain $\beta = -10$. In our experiment we observe that the two chaotic systems synchronize when $R_i \leq 5k \si{\ohm}$, i.e. $\beta = -R/R_i \leq -9.4$. Figure \ref{synch}(b) shows experimentally measured result obtained on an oscilloscope when $R_i = 5k \Omega$. The two signals captured on the oscilloscope are the error signal $e_1(t)$ (top) which converges to a small value when the controller is turned on using a Texas Instruments CD4066B electronic switch (bottom signal in Figure \ref{synch}(b) is the switch control signal).  The simulated time constant for the decay in error signal $e_1$, computed by fitting an exponential function to the envelope of the signal in Figure \ref{synch}(a) is $\tau_{sim}=5 \times RC=23.5\mu s$. The experimentally measured time constant for the decay in error signal $e_1(t)$ upon turning on the controller is $\tau_{expt}=300\mu s$. Figures \ref{synch} and \ref{synchxy} show the signals $x_1$ and $y_1$ in unsynchronized and synchronized states as observed on the oscilloscope.

\textcolor{blue}{}

\begin{figure}[!htb]
	\begin{center}
		\includegraphics[width=0.8\columnwidth]{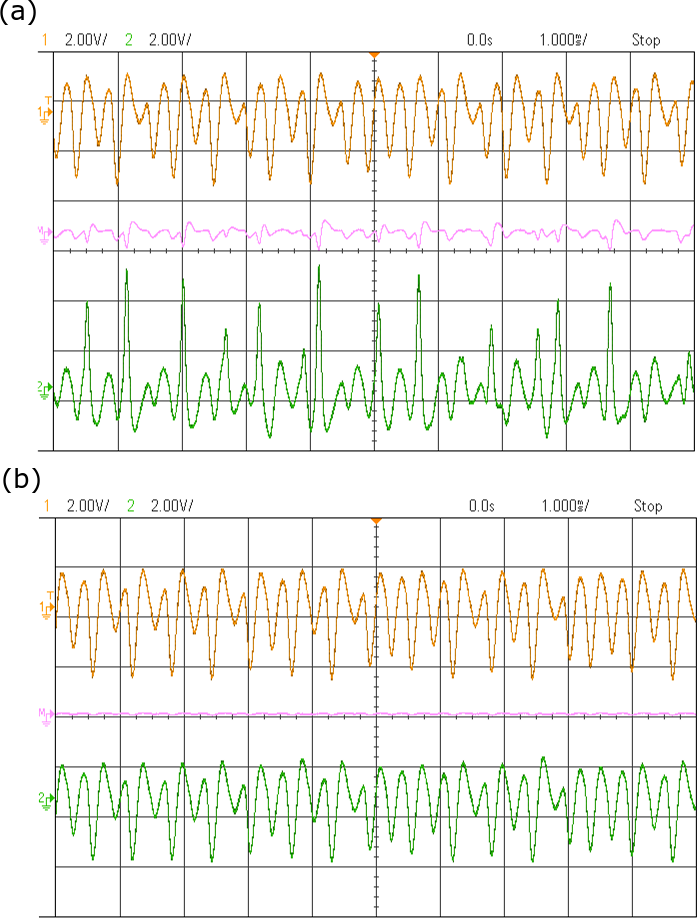}
	\end{center}
	\caption{Signals $x_1$ (master oscillator, upper trace), $y_1$ (slave oscillator, bottom trace) observed on an oscilloscope configured to display in time domain. The difference between the two signals is computed and displayed on the oscilloscope (middle trace). (a) When the two oscillators are not synchronized, the difference is non-zero. (b) The difference between $x_1$ and $y_1$ is very small, and the two traces look identical when the oscillators are synchronized.}
	\label{syncht}
\end{figure}

\section{Conclusion}

While synchronization of chaotic oscillator circuits has been demonstrated through several methods largely in the previous three decades, we present a method that utilizes a linear controller implemented using only one state signal from each oscillator circuit. This simultaneously makes the controller implementation extremely simple in an electronic circuit, and also enables cryptography applications wherein the unused state signals can be used for message encryption \cite{ref13a}. We also present a method to design a robust controller to achieve synchronization by analyzing the non-linear chaotic system as a multi-linear mode system and present a design methodology for the linear controller using root locus technique for ensuring stability. The analysis in this work and the method presented was developed specifically for the non-linearity in the oscillator circuit chosen for analysis in this work, and our future work will focus on developing a generalized design methodology and necessary and sufficient conditions for stability of any arbitrary multi-linear mode system, and exploring extending this result to a network of oscillators. 


\end{document}